\documentclass[letterpaper, 11 pt]{article} 
\usepackage{amsmath,amsfonts,amssymb,color}
\usepackage{mathtools}
\usepackage{dsfont}
\usepackage[dvipsnames]{xcolor}
\definecolor{mygreen}{rgb}{0.0, 0.5, 0.0}
\definecolor{winered}{rgb}{0.8,0,0}
\definecolor{myblue}{rgb}{0,0,0.8}
\usepackage{tikz}
\usepackage{amsthm}
\usepackage{empheq}
\newtheorem{theorem}{Theorem}

\newtheorem{definition}[theorem]{Definition}
\newtheorem{lemma}[theorem]{Lemma}
\newtheorem{proposition}[theorem]{Proposition}

\newtheorem{remark}[theorem]{Remark}

\DeclareMathOperator*{\argmax}{\arg\!\max}
\DeclareMathOperator*{\argmin}{\arg\!\min}
\DeclarePairedDelimiterX{\abs}[1]{\lvert}{\rvert}{#1}

\usepackage{graphicx}
\usepackage{subcaption}
\usepackage{wrapfig}
\usepackage{hyperref}
\usepackage{epsfig}
\usepackage{array}
\usepackage{multirow}
\usepackage{epstopdf}
\usepackage{tikz}
\usepackage{relsize}
\usetikzlibrary{shapes,arrows}
\hypersetup{
    colorlinks=true,
    linkcolor={winered},
    citecolor={mygreen}
}
\usepackage{cite}
\usepackage{enumitem}

\usepackage{xspace}
\newcommand\blfootnote[1]{%
  \begingroup
  \renewcommand\thefootnote{}\footnote{#1}%
  \addtocounter{footnote}{-1}%
  \endgroup
}

\usepackage[ruled]{algorithm2e}%
\usepackage[margin=1in]{geometry}
\title{Sensor Scheduling in Intrusion Detection Games \\with Uncertain Payoffs}
\author{Jayanth Bhargav, Shreyas Sundaram and Mahsa Ghasemi}
\date{}
\begin{document}
\maketitle
\begin{abstract}
We study the problem of sensor scheduling for an intrusion detection task. We model this as a 2-player zero-sum game over a graph, where the defender (Player 1) seeks to identify the optimal strategy for scheduling sensor orientations to minimize the probability of missed detection at minimal cost, while the intruder (Player 2) aims to identify the optimal path selection strategy to maximize missed detection probability at minimal cost. The defender’s strategy space grows exponentially with the number of sensors, making direct computation of the Nash Equilibrium (NE) strategies computationally expensive. To tackle this, we propose a distributed variant of the Weighted Majority algorithm that exploits the structure of the game's payoff matrix, enabling efficient computation of the NE strategies with provable convergence guarantees. Next, we consider a more challenging scenario where the defender lacks knowledge of the true sensor models and, consequently, the game's payoff matrix. For this setting, we develop online learning algorithms that leverage bandit feedback from sensors to estimate the NE strategies. By building on existing results from perturbation theory and online learning in matrix games, we derive high-probability order-optimal regret bounds for our algorithms. Finally, through simulations, we demonstrate the empirical performance of our proposed algorithms in both known and unknown payoff scenarios. \blfootnote{This material is based upon work supported by the Office of Naval Research (ONR) via Contract No. N00014-23-C-1016 and under subcontract to Saab, Inc. as part of the TSUNOMI project. Any opinions, findings and conclusions or recommendations expressed in this material are those of the author(s) and do not necessarily reflect the views of ONR, the U.S. Government, or Saab, Inc. \\
The authors are with the Elmore Family School of Electrical and Computer Engineering, Purdue University, West Lafayette IN 47907 USA. Email addresses: \{\tt jbhargav, \tt sundara2, \tt mahsa\}\tt @purdue.edu}
\end{abstract}

\textbf{Keywords:}
Online Learning, Matrix Games, Sensor Scheduling, Finite-Sample Convergence

\section{Introduction}
In the modern landscape of security and surveillance, efficient deployment and scheduling of sensors is critical in ensuring comprehensive monitoring and intrusion detection \cite{guvensan2011coverage,murray2007coverage}. In managing sensing resources, the challenge involves not only their placement \cite{dhillon2003sensor,bhargav2023complexity,jourdan2008optimal}, but also dynamically scheduling their orientations to maximize detection efficiency while minimizing resource usage and operational costs \cite{osais2009sensor}. Traditionally, sensor scheduling has been approached through various optimization techniques, yet these methods often fall short in adversarial settings where an intelligent intruder (non-oblivious)  can actively attempt to evade detection. In such scenarios, game theory offers a robust framework for modeling and solving the intricate interactions between the defending system and the intruding adversary \cite{pirani2021game,an2017stackelberg}. Furthermore, the defending system may not have accurate knowledge of sensor performance due to the inherent complexity of modeling certain sensors, such as cameras, or the lack of prior knowledge about detection probabilities. These uncertainties lead to discrepancies between the simulated and real-world game dynamics, requiring the defending system to learn the true game online, and adaptively refine strategies using feedback from sensors to ensure effective detection and maintain performance despite the model uncertainties.

\subsection{Related Work}
In this section, we review existing works on security games, methods for solving large-scale games, and online learning in games, highlighting how these approaches relate to our contributions.

\noindent \textbf{Security Games:} Many works in the literature have modeled security resource allocation tasks using the Stackelberg game framework \cite{sinha2018stackelberg,korzhyk2010complexity}. In \cite{paruchuri2008playing}, the authors model a patrolling task to protect airports from attackers as a Stackelberg game, where the defender commits to a mixed strategy for allocating patrolling resources and the attacker best-responds with a pure attack strategy. However, we consider the setting where both the players commit to mixed-strategies, and operate under concurrent decision-making conditions. Since neither of the players can anticipate or respond to the other’s exact strategy in real-time, the Stackelberg and Nash equilibria are equivalent in our game setting \cite{korzhyk2011stackelberg}. 

\noindent \textbf{Solving Large Games: }A line of work focuses on developing efficient techniques for approximately solving games with large strategy spaces \cite{ganzfried2015endgame,lipton2003playing,sandholm2015abstraction}. \cite{li2020structure} introduce an iterative structure-learning approach to search for approximate solutions of many-player games. In \cite{li2023decision} and \cite{brown2019deep}, the authors study team-adversary games with combinatorial action spaces and propose a counterfactual regret minimization framework for learning individual strategies. Along the lines of these works, we propose a distributed variant of the well-known Weighted Majority algorithm (proposed in \cite{littlestone1994weighted}) which leverages the structure in the game's payoff function, to efficiently and iteratively estimate the NE, and establish convergence guarantees.

\noindent \textbf{Online Learning in Games: } In \cite{fiez2021online}, the authors study online learning in periodic zero-sum games and establish guarantees for convergence to equilibrium strategies. The authors of \cite{bailey2019fast} study online learning in $2\times 2$ zero-sum games and provide convergence guarantees for gradient descent dynamics. However, for a general class of zero-sum games, the dynamics of the equilibrium strategies generated by the learning process can be very intricate and instances may fail to converge \cite{andrade2021learning,cheung2019vortices}. Two closest papers to our work on learning in matrix games with bandit feedback are: \cite{o2021matrix} - where the authors propose a UCB style algorithm for no-regret learning in matrix games; and \cite{chen2024last} - where the authors present an adaptation of online mirror descent algorithm with last-iterate convergence rates. The key distinction of our work is the application to the intrusion detection setting, where we exploit the combinatorial structure of the game matrix arising from sensor and path configurations to establish tighter regret bounds and faster convergence guarantees.

\subsection{Contributions}
We begin by modeling the sensor scheduling task as a zero-sum matrix game. First, we propose a distributed variant of the Weighted Majority algorithm that leverages the structure of the game's payoff matrix to efficiently compute NE strategies, even in exponentially large strategy spaces, with provable convergence guarantees. Second, recognizing that the defender may lack accurate knowledge of the true sensor models and hence the game's payoff matrix, we develop online learning algorithms that rely on bandit feedback from the sensors to adaptively estimate the game matrix and refine the defender's sensor scheduling strategies. By building on results from perturbation theory and online learning in matrix games, we derive high-probability order-optimal regret bounds for these algorithms. Finally, we demonstrate the empirical performance of our methods through numerical simulations, for both known and unknown payoff scenarios.
\vspace{-3mm}
\section{Problem Formulation}
\label{sec:prob_form}
We model the game environment as a graph $G = (N,E)$, where the set of nodes $N$ denotes the regions of an environment, and the set of edges $E$ denotes the possible transitions for the intruder between the regions. Let $\mathcal{S} = \{s_1, \hdots, s_p\}$, with $\abs{\mathcal{S}} = p$, denote the set of sensors deployed on a subset of nodes in the graph. Each sensor has $d$ possible orientations, $\Theta = \{ \theta_1, \hdots, \theta_d\}$, with $\abs{\Theta} = d$. Let $c^q_{k} \in \mathbb{R}_{\ge 0}$ be the cost of orienting the sensor $s_q \in \mathcal{S}$ in the direction $\theta_k \in \Theta$. Each sensor covers a subset of nodes in the graph depending on its orientation, which may not be restricted to its neighboring nodes.  Let $I = \{i_1, i_2, \hdots, i_m \}, m = |I|$, denote the set of joint strategies of the defender, i.e., a set of all possible sensor orientations, and let $\{ c_1, c_2, \hdots, c_m \}$ be the costs associated with each joint strategy. The cost $c_i$ of a joint strategy $i \in I$ is the sum of the individual costs associated with each sensor orientation in that strategy, i.e., $c_i = \sum_{q = 1}^p c^q_{k_{q,i}}$, where $k_{q,i}$ is the orientation of sensor $s_q$ in the joint strategy $i$.  We note that the number of joint strategies is $m = \abs{\Theta}^{\abs{\mathcal{S}}}$, which is exponential in the number of sensors. We assume that the intruder starts at a source node $S$ and aims to reach a terminal node $T$ in the graph, and has a finite set of possible paths which it can take. We denote this by $J = \{ j_1, j_2, \hdots, j_n\}, n=|J|$,  which is the set of pure strategies (paths) of the intruder, and let $\{ r_1, r_2, \hdots, r_n \}, r_j \in \mathbb{R}_{\ge 0}$, be the associated costs of the respective paths. Let $X$ and $Y$ denote the set of mixed-strategies for the defender and intruder, respectively. We have the following:
\begin{equation}\label{eq:strategy_space}
    X = \left\{ x \in \mathbb{R}^m ; \sum_{i\in I} x_i = 1, x \geq 0  \right \};Y = \left\{ y \in \mathbb{R}^n ; \sum_{j\in J} y_j = 1, y \geq 0  \right \}.
\end{equation}
Each sensor $s_q \in \mathcal{S}$ has a probability of detection $p_{detect,q}$. Specifically, if the intruder traverses through a node which is covered by a sensor $s_q$, it will be detected by that sensor with a probability $p_{detect,q}$, and will go undetected by that sensor with probability $(1-p_{detect,q})$. We now make the following assumption on the sensor models.\\

\textbf{Assumption 1:} The sensors are imperfect with $p_{detect,q} \in [p_{min,q}, p_{max,q}]$, where $0<p_{min,q} < p_{max,q} < 1, q = 1,...,p$. Furthermore, a detection event from a sensor $s_q$ is independent of detection events from other sensors in $\mathcal{S}\setminus\{s_q\}$.\\

 The goal of the defender is to minimize the probability of not detecting the intruder. If a certain node is covered by multiple sensors and the intruder has traversed that node, then the overall probability of missed detection at that node is the product of the probability of missed detection of all the sensors covering that node. For a scheduling strategy $i\in I$ and a path $j \in J$, let $V^q_{ij}$ be the number of nodes that are covered by the sensor $s_q$ in the orientation strategy $i \in I$ and contained in the intruder's path $j \in J$. The overall probability of missed detection is given by:
\begin{equation}
\label{eq:pmiss_prod}
    p_{miss}(i,j) = \prod_{q=1}^p (1-p_{detect,q})^{ V^q_{ij}}.
\end{equation}
We consider the additive log-likelihood representation for the probability of missed detection  (as multiplicative operations in \eqref{eq:pmiss_prod}  can lead to underflow and numerical instability), given by
\begin{equation}
    \log(p_{miss}(i,j)) =  \sum_{q=1}^p V^q_{ij}  \log (1-p_{detect,q}).
\end{equation}

Let $A,B \in \mathbb{R}^{m \times n}$ denote the payoff matrices for the defender and intruder, respectively. The payoff values for a pure strategy pair $(i,j)$ for the defender and intruder are given by $A_{ij} =  \log(p_{miss}(i,j)) + c_i$ and $B_{ij} =   -\log(p_{miss}(i,j)) + r_j$, respectively. The  defender plays $x \in X$, the  intruder plays $y \in Y$, and they receive payoffs $x^{\top}Ay$ and, $x^{\top}By$ respectively. 
 The defender aims to find the mixed strategy $x^*$ that minimizes the expected value of its payoff, given the mixed strategy $y$ of the intruder, i.e., the defender aims to solve the following optimization problem:
\vspace{-1mm}
\begin{equation}
\label{eq:defender_opt}
    x^* = \argmin_{x \in X} x^{\top}Ay.
\end{equation}
Similarly, given the mixed strategy $x$ of the defender, the intruder aims to solve the following:
\begin{equation}
\label{eq:intruder_opt}
    y^* = \argmin_{y \in Y} x^{\top}By.
\end{equation}
This will form a  bimatrix game which we denote by the tuple: $\mathcal{G} = \{A,B\}$. Note that $\mathcal{G}$ is a non-zero sum game, since $A+B \neq 0$ in general. A bimatrix game $\mathcal{G} := \{A,B\}$ is zero-sum if $A+B=0$. The solution to a zero-sum game, i.e., the Nash equilibrium strategies $(x^*,y^*)$, can be obtained by solving the nested min-max optimization problem below: 
 \vspace{-1mm}
\begin{equation}
\label{eq:minmaxgame}
   (x^*,y^*) \in \argmin_{x \in X} \argmax_{y\in Y} x^{\top}Ay.
\end{equation}
\begin{definition}[Nash Equilibrium]
    A pair of mixed-strategies $(x^*,y^*)$ is a \textit{Nash Equilibrium} of the zero-sum game $A$ if $ x^{{* \top}}A y^* \leq x^{\top}Ay^*$ and $ x^{{* \top}}A y^* \geq x^{{* \top}}Ay, \hspace{2pt} \forall x \in X, \hspace{1pt} \forall y \in Y$.
\end{definition}
Consider the following $\mathcal{G}' := \{A',B'\}$, where matrices $A'$ and $B'$ are such that 
\begin{equation}
\label{eq:eq_game_def}
\begin{aligned}
    A'_{ij} &= A_{ij} - r_j; \quad  B'_{ij} &= B_{ij} - c_i.
\end{aligned}
\end{equation}
Based on the results presented in Section 2.1 of \cite{kannan2010games}, we have the following result that characterizes the equilibrium property of $\mathcal{G}'$. All detailed proofs supporting our theoretical results are deferred to Appendix \ref{app:proof}.
\begin{proposition}
\label{prop:1}
    The game $\mathcal{G}' := \{ A',B'\}$ is zero-sum and has the same set of Nash Equilibria as $\mathcal{G}$.
\end{proposition}
  We consider the setting in which the defender has knowledge of $c_i$'s and $r_j$'s for all $i \in I$ and $j\in J$. However, the defender may not know the sensor detection probabilities $p_{detect}$, consequently the game payoff matrix $A$, accurately. Proposition \ref{prop:1} also implies that an $\epsilon$-NE\footnote{A pair of strategies $(x',y')$ is a $\epsilon$-NE of $A$ if $ x'^{{ \top}}A y' \leq x'^{ \top}Ay+\epsilon$ and $ x'^{{ \top}}A y' \geq x^{{ \top}}Ay' - \epsilon, \hspace{2pt} \forall x \in X, \hspace{1pt} \forall y \in Y$.} of $(A',B')$ is also an $\epsilon$-NE for the original game $(A,B)$. With a slight abuse of notation, we will denote the modified game (which is zero-sum) by $\mathcal{G}$, which can be completely specified by the payoff matrix $A$, where $A_{ij} = \log(p_{miss}(i,j)) +c_i-r_j$.

\begin{remark}
    We assume that the entries of $A$ are bounded in $[0,1]$. If this is not the case, we normalize $A$ using its maximum and minimum values, $A_{min}=\min_{i,j} A_{ij}$ and $A_{max}=\max_{i,j} A_{ij}$, respectively. This normalization preserves the Nash equilibrium strategies, as scaling and shifting the payoffs affect only the game's value, not the equilibrium strategies themselves. Additionally, we note that an $\epsilon$-NE of the normalized game is an $\epsilon'$-NE of the un-normalized game, where $\epsilon' =  (A_{max}-A_{min}) \epsilon$. Therefore, it is sufficient to restrict all analysis to the normalized zero-sum game, as it fully captures the essential properties and results of the original game.
\end{remark}

\section{Distributed Weighted Majority Algorithm}

In this section, we propose a fast and scalable algorithm which leverages the structure of the game matrix $A$ for efficiently solving for the NE strategies. For a moderate number of strategies, the min-max optimization problem in \eqref{eq:minmaxgame} can be solved to compute the NE strategies using a bi-level Linear Programming (LP) formulation \cite{nehme2009two}. However, for the sensor scheduling problem considered in this paper, the defender has an exponentially sized set of scheduling strategies. For example, if $d=\lvert\Theta \rvert = 4$ (i.e., each sensor has 4 possible orientations) and $\lvert\mathcal{S}\rvert=10$ (i.e., there are $10$ sensors), then $\lvert I\rvert = 4^{10}=1,048,576$.  Solving for the exact NE using an LP can become computationally intractable due to a large number of variables and constraints. To this end, we present the Distributed Weighted Majority (DWM) algorithm (Algorithm \ref{alg:dwm}), a variant of the \textit{Weighted Majority (WM)} algorithm originally developed by \cite{littlestone1994weighted}. The DWM algorithm leverages the structure of the game to perform multiplicative weight updates locally for each sensor, which will then be aggregated to estimate the joint scheduling strategy for the defender.

First, we define the set of sub-game payoff matrices $\mathcal{A} = \{ A^1, A^2, \hdots, A^p \}$, where $A^q \in \mathbb{R}^{d \times n}$ corresponds to sensor $s_q \in \mathcal{S}$, given by
\begin{equation}
    \label{eq:small_A}
    A^q_{kj} =
     V^q_{kj} \times \log(1-p_{detect,q}) + c_k^q,
\end{equation}
where $V^q_{kj}$ contains all the nodes that are covered by the sensor $s_q$ oriented in the direction $\theta_k$ and contained in the intruder path $j$. At each iteration $t$ and for each sensor $s_q \in \mathcal{S}$, we maintain a set of weights $\sigma^q_t \in \mathbb{R}^{d}$, where $d = |\Theta|$, which are proportional to the likelihood of selecting different orientations for the sensor. The joint sensor scheduling strategy $i \in I$ can be denoted as $i = [k_{1,i}, k_{2,i}, \hdots, k_{p,i}]$, where $k_{q,i} \in \{ \theta_1, \hdots, \theta_d \}$ is the orientation of the sensor $s_q \in \mathcal{S}$ in the joint strategy $i \in I$. We update the weights for each joint strategy $i \in I$ as follows:
\begin{align}
\label{eq:strat_con}
    w_t(i) = \prod_{q = 1}^{p} \sigma^q_t(k_{q,i}),
\end{align}
where $\sigma^q_t(k_{q,i})$ is the weight assigned to the orientation $\theta_k$ for the sensor $s_q$ in the joint strategy $i\in I$.
At each iteration $t$, the weights $\sigma^q_t(k_{q,i})$ are updated using the multiplicative rule in \eqref{eq:dist_weight} using a learning rate $\beta \in (0,1)$ and the loss per sensor $(s_q \in \mathcal{S})$ for each individual strategy $\theta_k \in \Theta$, denoted as $\ell^q_t(\theta_k)$, given by
\begin{equation}
    \label{eq:agentloss}
    \ell^q_t(\theta_k) = \sum_{j \in J} A^q_{kj}y_t(j) - \frac{1} {p}\sum_{j\in J} r_j y_t(j).
\end{equation}

\begin{algorithm}[!h]
\caption{ Distributed Weighted Majority (DWM) Algorithm}\label{alg:dwm}
\KwData{Intrusion Detection Game $\mathcal{G}$, Parameter $\beta \in (0,1)$, Number of Iterations $T$}
\KwResult  {Mixed-strategies: $\{x_1, \hdots, x_T\}, \{y_1, \hdots, y_T\}$}
\textbf{Initialize:} $\sigma_0^q(\theta_k) = 1/{|\Theta|}, k = 1,...,|\Theta|$; $w_0(i) = 1/|I|, i = 1,...,m$\\
\For {$t = 1, \hdots, T$}{
\hspace{1mm} Compute defender's mixed-strategy $x_t(i) = \frac{w_t(i)}{\sum_{i'\in I} w_t(i')}, \forall i \in I$\\
\noindent  Estimate intruder's mixed-strategy $y_t = \max_{y \in Y} x_t^{\top}Ay$ \\
\noindent Compute per sensor loss $\ell^q_t(\cdot)$ as in Equation \eqref{eq:agentloss}\\
\noindent Update weights for each sensor
\begin{equation}
    \label{eq:dist_weight}
    \sigma_{t+1}^q(k_{q,i}) = \sigma_{t}^q(k_{q,i}) \beta^{\ell_t^q(\theta_k) }; \quad k = 1,\hdots,|\Theta|, q = 1,...,p
\end{equation}
\noindent Update joint strategy weights:
$w_{t+1}(i) = \prod_{q = 1}^{p} \sigma^q_{t+1}(k_{q,i}), \forall i \in I$
}
\end{algorithm}
Based on the results in Sections 2.4 and 2.5 of \cite{freund1996game}, we have the following theoretical guarantees for Algorithm \ref{alg:dwm}.
\begin{theorem}
\label{thm:dwm}
    Let $\{ x_1, \hdots, x_T \}$ and $\{ y_1, \hdots, y_T \}$ be the mixed strategies returned by Algorithm \ref{alg:dwm} after $T$ iterations with $  \beta = \left(1+ \sqrt{\frac{2 |\mathcal{S}|ln|\Theta|}{\Tilde{L}}}\right)^{-1},$ where $ \Tilde{L} \ge \min_{i \in I} \sum_{t=1}^{T} \sum_{q=1}^p \ell^q_t(\theta_k) $. Let $\bar{x} = \frac{1}{T} \sum_{t=1}^{T}x_t$ and $\bar{y} = \frac{1}{T} \sum_{t=1}^{T}y_t$. Then, $\Bar{x}$ and $\Bar{y}$ are valid mixed-strategies for the two players, respectively. The pair $(\Bar{x},\Bar{y})$ approximates the game's Nash Equilibrium value within $\epsilon_T$, given by

    \begin{equation}
    \label{eq:nashgap}
        \epsilon_T = \frac{\sqrt{2 \Tilde{L} |\mathcal{S}|ln|\Theta|}}{T} + \frac{|\mathcal{S}|ln|\Theta|}{T}.
    \end{equation}
\end{theorem}
$\Tilde{L}$ in Theorem \ref{thm:dwm} is an upper bound on the sum of losses over $T$ iterations, which can be naively bounded by $T$. For a desired approximation $\epsilon$, running Algorithm \ref{alg:dwm} for $T$ iterations (where $T$ can be computed using \eqref{eq:nashgap}), with the specified learning rate $\beta$, will yield an $\epsilon$-NE strategy $(\bar{x},\bar{y})$.
\begin{remark}
    The DWM algorithm (Algorithm \ref{alg:dwm}) exploits the additive structure of the game's payoff function, achieving a substantial computational advantage by performing only $\mathcal{O}(|\mathcal{S}||\Theta|)$ multiplicative updates in parallel (i.e., $\mathcal{O}(|\Theta|)$ updates for each sensor $s \in \mathcal{S}$ in parallel), to estimate the NE strategy for the defender. In contrast, directly applying the WM algorithm requires $\mathcal{O}(|\Theta|^{|\mathcal{S}|})$ updates, which cannot be parallelized, making DWM far more efficient for large-scale instances. Furthermore, in Line 2 of Algorithm \ref{alg:dwm}, we estimate the intruder’s mixed strategy $y_t$	
  by solving a standard LP, as the intruder’s strategy space is significantly smaller than that of the defender $(n << m)$. However, if the intruder’s strategy space is also large, the multiplicative weight update rule can be applied to update $y_t$. This approach, however, results in a significantly slower convergence rate compared to a single player using the multiplicative weight update.
\end{remark}

\section{Online Learning and Adaptation for Unknown Sensor Models}
In this section, we extend our study of the sensor scheduling problem to scenarios where the defender lacks knowledge of the true sensor models and, consequently, the
game’s payoff matrix. We present online learning algorithms with performance guarantees for both homogenous and heterogeneous sensor settings, for refining the defender's strategies based on bandit feedback.  

Since the defender does not have accurate knowledge of the game's payoff, it maintains an estimate of the game matrix, which is updated iteratively using feedback from sensor observations. At each time step $t$, the defender computes a mixed strategy $x_t \in X$ for scheduling sensors, based on its current estimate of the game matrix. The defender and intruder play pure strategies $(i_t,j_t)$, sampled from their mixed-strategy distributions $(x_t,y_t)$ respectively. The defender receives feedback from the sensors monitoring the intruder's path, indicating whether or not the intruder was detected. 
\newpage
We will first consider the setting with homogenous sensors, where the game matrix $A$ can be parametrized by a single (unknown) Bernoulli distribution with parameter $p_{detect}$. As a result, the game matrix can now be defined as $A_{ij} = V_{ij} \log{(1-p_{detect})} + c_i - r_j$, where $V_{ij} = \sum_{q=1}^p V^q_{ij}$.  

We define $\mathcal{F}_t = ((i_1, j_1), \hdots,(i_t, j_t)) $ to be the sequence of observations available to the player at round $t$. Two aspects of this problem distinguish it from other setups considered in the literature (\cite{cheung2019vortices,andrade2021learning,o2021matrix}). Firstly, the defender receives the pure strategy played by the intruder $(j_t)$ as feedback (as opposed to $y_t$), and secondly, the intruder does not have accurate knowledge of the game payoff function/matrix. Additionally, our setting is different from the one considered in \cite{o2021matrix}, where the players receive a noisy feedback of the payoff, and the noise process is assumed to be 1-sub-Gaussian with a zero mean.  At each time step $t$, the defender updates its estimate of $\hat{p}_{detect}^t$ of the true parameter $p_{detect}$  and estimates the game matrix $\hat{A}_t$ (see Algorithm \ref{alg:1}). Another key distinction from \cite{o2021matrix} is that our problem setting under homogenous sensors does not need complete exploration of the strategy spaces. We introduce a simpler algorithm that eliminates the need for Upper Confidence Bound (UCB) estimates for each entry of the 
$A$ matrix, as the exploration-driven optimism framework is not required in this scenario. Consequently, we establish tighter regret bounds that do not depend on the strategy spaces, i.e., ($\tilde{\mathcal{O}}(\sqrt{T})$ instead of $\tilde{\mathcal{O}}(\sqrt{|I||J|T})$ ).

\begin{algorithm}[!h]

\caption{ Online Learning and Adaptation of Strategies for Homogenous Sensors}\label{alg:1}

$\text{Feedback samples: }\mathcal{H} = \emptyset $\\
\For {$t = 1, \hdots, T$}{
Compute sample average $\hat{p}_{detect}^t$ using samples in $\mathcal{H}$\\
Clipping estimator: $\hat{p}_{detect,l}^t = \min (  \max(\hat{p}^t_{detect,l},p_{min,l}),p_{max,l}); l = 1,...p$\\
Solve the estimated game $\hat{A}_t$, where $\hat{A}_t(i,j) = V_{ij}\log(1-\hat{p}_{detect}^t) + c_i - r_j$:
\begin{equation} \label{eq:solve_game}
    \displaystyle  (x^*_t,y^*_t) \in \displaystyle \argmin_{x\in X} \argmax_{y \in Y} x^{\top}\hat{A}_ty \quad \text{(using Algorithm \ref{alg:dwm})}\end{equation}
Play pure strategy $i_t \sim x^*_t$ and observe intruder's pure strategy $j_t \sim y_t$ \\
Collect $V_{i_t,j_t}$ samples from sensor feedback into $\mathcal{H}$ }

\end{algorithm}
In Algorithm \ref{alg:1}, we use a clipped estimator for the parameter $\hat{p}^t_{detect}$, which keeps the estimate within a feasible range  $[p_{min},p_{max}]$ ; this ensures stability in small sample sizes without affecting long-term convergence to the true parameter $p_{detect}$. Furthermore, in each iteration of Algorithm \ref{alg:1}, the estimated game (as in \eqref{eq:solve_game}) can be efficiently solved for NE strategies using Algorithm \ref{alg:dwm}.

We will perform analysis from the perspective of a single player, i.e., the defender, who does not control the actions of the opponent (i.e., the intruder). Let $(x^{\dagger},y^{\dagger})$ denote a pair of NE strategies of the game $\mathcal{G}$ (with payoff matrix $A$), and $V^*_A = x^{\dagger \top} A y^{\dagger}$ be the value of the true game. In order to analyze the performance of the online learning algorithm, we consider the individual regret for the defender, i.e., the difference in expected cumulative payoffs relative to the value of the game:
\begin{equation}
\label{eq:regret}
     \mathcal{R}_T = \sum_{t = 1}^{T} \displaystyle \left( x_t^{\top} A y_t - V^*_A \right).
\end{equation}


Building on the results from perturbation theory in zero-sum games presented in \cite{lipton2006stability} we present the following result, characterizing the bound on the cumulative regret in \eqref{eq:regret}.  
\newpage
\begin{theorem}
\label{thm:regret}
    Under Assumption 1, for a specified $\alpha \in (0,1)$, with probability at least $1-\alpha$, Algorithm \ref{alg:1} has the following guarantees for the cumulative regret  (as in \eqref{eq:regret}):
    \begin{enumerate}
        \item $\displaystyle \mathcal{R}_T  \leq \frac{5 \sqrt{2} V_{max}}{(1-p_{max})} \sqrt{T \log{\frac{2T}{\alpha}}}; y_t = y^{\dagger} \hspace{2pt} \text{(i.e., $y_t$ is any NE of $\mathcal{G}$)}; $
        \item $\displaystyle \mathcal{R}_T  \leq \frac{2 \sqrt{2} V_{max}}{(1-p_{max})} \sqrt{T \log{\frac{2T}{\alpha}}}; y_t \neq y^{\dagger} \hspace{2pt} \text{(i.e., $y_t$ is not a NE of $\mathcal{G}$)},$
   \end{enumerate}
    where $\displaystyle V_{max} = \max_{i,j} V_{ij}$ and $p_{max}$ is the maximum probability of detection of the sensors. 
\end{theorem}

\begin{remark}
   In \cite{cai2024uncoupled}, the authors propose an online learning algorithm which is an adaptation of \textsc{Exp3-IX}, where they assume the existence of mixed-strategy Nash equilibrium with full support. Particularly, at each iteration $t$ of their algorithm, they restrict the space of NE strategies to $\Omega_{t} = \{ x \in X: x_a \geq \frac{1}{|I|t^2}, \forall a \in I \}$. However, many zero-sum games are not guaranteed to have a mixed-strategy Nash equilibrium with full support. In our algorithm, we do not restrict the space of NE strategies for the players. Additionally, our results presented in Theorem \ref{thm:regret} provide tighter regret guarantees $\Tilde{O}(T^{-\frac{1}{2}})$ compared to $\Tilde{O}(T^{-\frac{1}{8}})$ presented in \cite{cai2024uncoupled}. 
\end{remark}

We will now consider the setting with heterogeneous sensors, i.e., each sensor has a different probability of detection. In this setting, the game's payoff matrix $A$ can be parameterized by a set of $|\mathcal{S}|$ Bernoulli distributions, with parameters $\{ p_{detect,1}, \hdots, p_{detect,p} \}, p=|\mathcal{S}|$, each corresponding to a sensor $s_l \in \mathcal{S}$, $l = 1,\hdots,p$. The game's payoff matrix in this setting is given by: 
\begin{equation}
    \label{eq:hetero_A}
    A_{ij} = \sum_{l=1}^{p} V_{ij,l} \log(1-p_{detect,l}) +c_i - r_j,
\end{equation}
where $V_{ij,l}$ is the number of nodes covered by the sensor $l$ in the joint strategy $i \in I$ and contained in the intruder's path $j \in J$, and $c_i,r_j$ are sensor and path costs as specified before, respectively. Define $\max_{ij} V_{ij,l} = V_{max,l}$. We note that each entry of $A$ may not depend on all $|\mathcal{S}|$ Bernoulli parameters, as $V_{ij,l}$ can be zero for a subset of sensors in $\mathcal{S}$. This is due to the fact that every intruder path $j \in J$ will intersect nodes covered in the graph by only a subset of the sensors. Thus, unlike the homogenous sensors setting, we will need exploration of strategies of both the players to ensure better estimates of these Bernoulli parameters. 

We adapt the Upper Confidence Bound (UCB) algorithm for learning in matrix games proposed in \cite{o2021matrix}, by constructing the UCB estimate for the game matrix $A$ based on the confidence bounds of the Bernoulli parameters.  
We assume that $c_i$'s and $r_j$'s are known.

\begin{lemma}
    For the game matrix A defined in \eqref{eq:hetero_A}, the following holds with probability at least $1-\delta$,
\begin{equation}
\label{eq:UCB_A}
    A_{ij} \leq \bar{A}^t_{ij} + \sum_{l=1}^{p} \frac{V_{max,l}}{1-p_{max,l}} \sqrt{\frac{\log(\frac{2}{\delta})}{2 (1 \vee n^t_{ij,l})}} = \Tilde{A}^t_{ij}, \forall i,j
,\end{equation}
where $\bar{A}^t_{ij}$ is an estimate of $A_{ij}$ evaluated with sample averages $\hat{p}^t_{detect,l}$,  $n^t_{ij,l}$ is the number of times the intruder chooses the path $j$ which contains nodes covered by the sensor $l$ in the joint orientation strategy $i$ up to (but not including round $t$), and we use the notation $(1 \vee \cdot) = \max(1,\cdot)$.
\end{lemma}

Define $\bar{V}_{max} = \max_{l} V_{max,l}$ and $\bar{p}_{max} = \max_{l} p_{max,l}$.  We have the following result that characterizes the bound on the regret (Equation \eqref{eq:regret})  when Algorithm 1 in \cite{o2021matrix} is applied to the instance of the intrusion detection game with the UCB estimate $\Tilde{A}^t_{ij}$ as defined in Equation \eqref{eq:UCB_A}.

\begin{theorem}
\label{thm:ucb_regret}
     Let Assumption 1 hold and let $T \ge |\mathcal{S}||\Theta||J|\ge 2$, $\delta = 1/(2T^2|\mathcal{S}||\Theta||J|)$, $C = \bar{V}_{max}/(1-\bar{p}_{max})$. When Algorithm 1 in \cite{o2021matrix} is applied to the instance of the game $\mathcal{G}$ with the  UCB estimate $\Tilde{A}^t_{ij}$ (as in \eqref{eq:UCB_A}), the regret (as in \eqref{eq:regret}) is bounded as:
\vspace{-1mm}
    \begin{equation}
        \mathcal{R}_T \leq 1 + C |\mathcal{S}|\sqrt{2 |\Theta||J|T \log(4T^2|\mathcal{S}||\Theta||J|)} = \Tilde{O}(\sqrt{|\mathcal{S}|^2|\Theta||J|T}).
    \end{equation}
\end{theorem}

\begin{remark}
    Applying Algorithm 1 of \cite{o2021matrix} directly would result in a regret bound that depends exponentially on the number of sensors, i.e., $\Tilde{O}(\sqrt{|\Theta|^{|\mathcal{S}|}|J|T})$. We leverage the specific structure of the game to achieve significant improvements in the regret bound in Theorem \ref{thm:ucb_regret}, i.e., $\Tilde{O}(\sqrt{|\mathcal{S}|^2|\Theta||J|T})$,  which will lead to faster convergence to the NE strategy for the defender. Additionally, the parameter $\delta$, which governs the balance between exploration, regret bounds, and high-probability guarantees depend on $|\Theta||\mathcal{S}||J|$ instead of $|\Theta|^{|\mathcal{S}|}|J|$.  We present the adapted UCB algorithm (Algorithm \ref{alg:ucb}) and additional discussions in Appendix \ref{app:ucb}.
\end{remark}

\section{Experiments}
In this section, we evaluate the performance of the proposed algorithms through simulations.

\subsection{Empirical Evaluation of Distributed Weighted Majority Algorithm}
In this section, we present numerical evaluations to show the computational efficiency of DWM algorithm.  We consider the nine-room grid world environment as shown in Figure \ref{fig:fig1}(a). This is a $19 \times 19$ cell environment, which can be represented by a graph. Sensors are installed on the cells (resp. nodes in the graph) marked in red (and numbered), and all cells marked in black are obstacle cells. We set $\abs{\Theta}=4$, i.e., each sensor has $4$ possible orientations: \{ East, North, West, South\}. We vary the number of sensors $\abs{\mathcal{S}} \in \{2,3,4,5,6,7,8\}$ to generate multiple instances of the game. 
The defender's strategy space, given by $4^{\abs{\mathcal{S}}}$, scales exponentially with the number of sensors in $\mathcal{S}$. The intruder's strategy space is set to $\abs{J} = 50$, i.e., there are $50$ possible paths for the intruder (e.g., see Figure \ref{fig:fig1}(b)) from node $S$ (marked in yellow) to  node $T$ (marked in orange) in the grid. All simulations are run on a 2.6 GHz 6-Core Intel Core i7 machine. We set $\epsilon_T = 0.001$, i.e., we wish to compute a $\epsilon$-Nash Equilibrium for the game with $\epsilon = 0.001$. For each instance of the game, we compute the number of iterations $T$ based on $\abs{I}$ and $\epsilon_T$ using Equation \eqref{eq:nashgap} by setting $\Tilde{L} = T$. We solve for the NE strategies using three approaches: (i) a standard bi-level Linear Program (LP) (using Gurobi Optimization Solver), (ii) WM algorithm (as in \cite{littlestone1994weighted}) and (iii) DWM algorithm (Algorithm \ref{alg:dwm}) for $T$ iterations, and record the running times in Table \ref{tab:table1}. We observe that the WM and DWM algorithms solve the game to within a 0.001 approximation of the game value much faster than the LP. We observe that as the strategy space increases, the DWM algorithm outperforms the WM algorithm, with a significant reduction in running time. Due to limitations on the problem sizes in Gurobi Solver (academic license), we cannot solve LPs for problem instances with strategy spaces more than $2000$.
\vspace{-1mm}
\begin{table}[!h]
    \centering
    \begin{tabular}{ccccc}
    \hline
$|\mathcal{S}|$ & Strategy Spaces & Linear Program & Weighted Majority & Distributed Weighted Majority \\
       \hline 
       \hline
       $2$ & $16 \times 50$ & 3.22  & 2.73  & 2.68 \\
         $3$ &  $64 \times 50$ & 7.90  & 3.04  & 3.65 \\
        $4$ &  $256 \times 50$ & 24.88  & 6.09  & 4.28 \\
        $5$ &  $1024 \times 50$ & 75.72  & 12.15  & 6.13 \\
       $6$ &   $4096 \times 50$ & - & 35.51  & 7.88 \\
      $7$ &    $16,384 \times 50$ & - & 107.65  & 9.92 \\
        $8$ &  $65,536 \times 50 $ & - & 365.43  & 11.83 \\
         \hline
    \end{tabular}
    \caption{Comparison of running times (in seconds) of Distributed Weighted Majority Algorithm (Algorithm \ref{alg:dwm}) with standard Weighted Majority algorithm and Linear Program}
    \label{tab:table1}
\end{table}

    
    
\vspace{-2mm}
\subsection{Empirical Evaluation of Online Learning Algorithm}
\begin{figure*}[!ht]
    \centering
    \begin{subfigure}{0.27\textwidth}
        \centering
        \includegraphics[width=100pt]{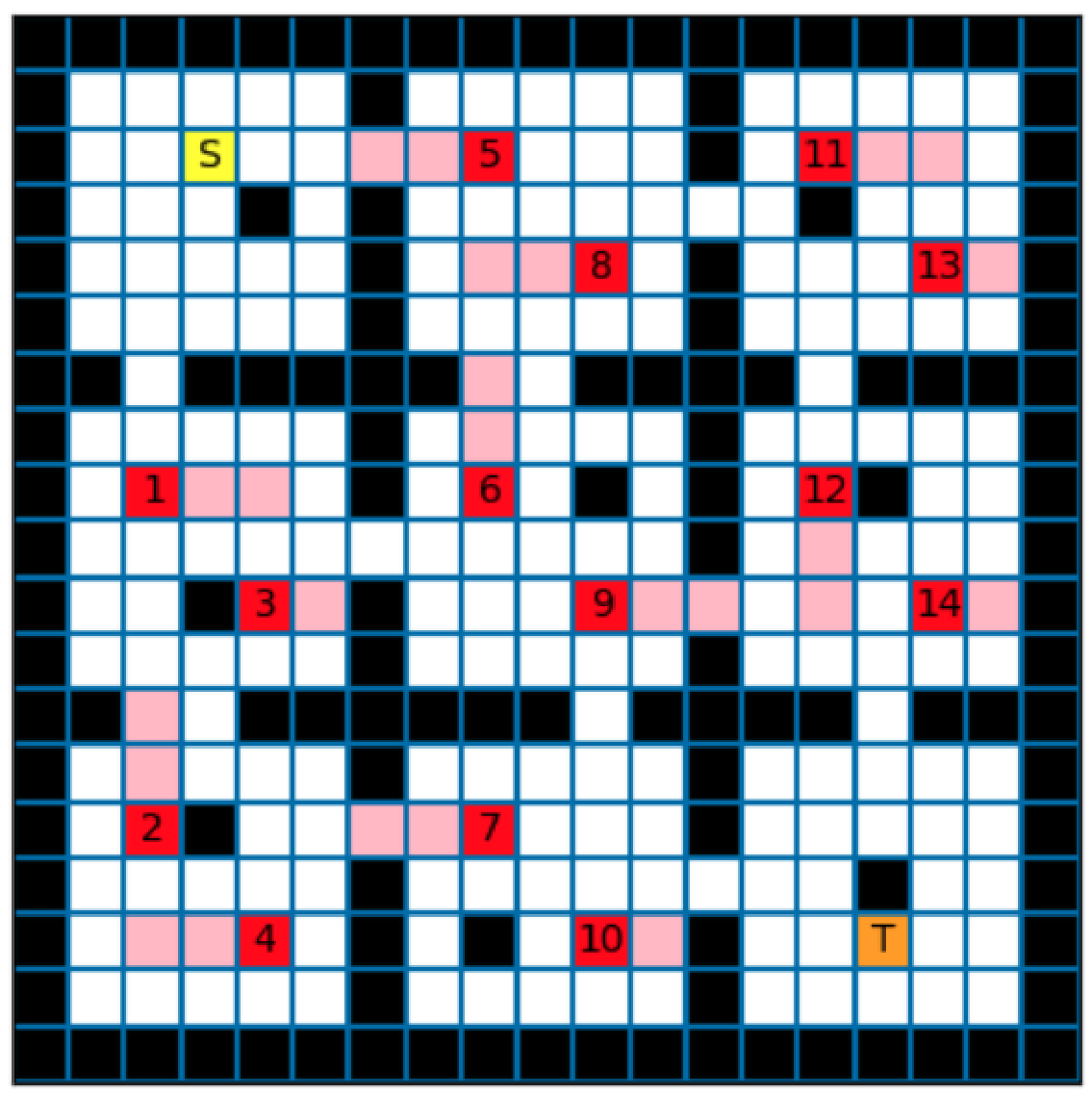}
        \par\small (a)
        \label{fig:a}
    \end{subfigure}
    \hspace{1mm}
    \begin{subfigure}{0.28\textwidth}
        \centering
        \includegraphics[width=100pt]{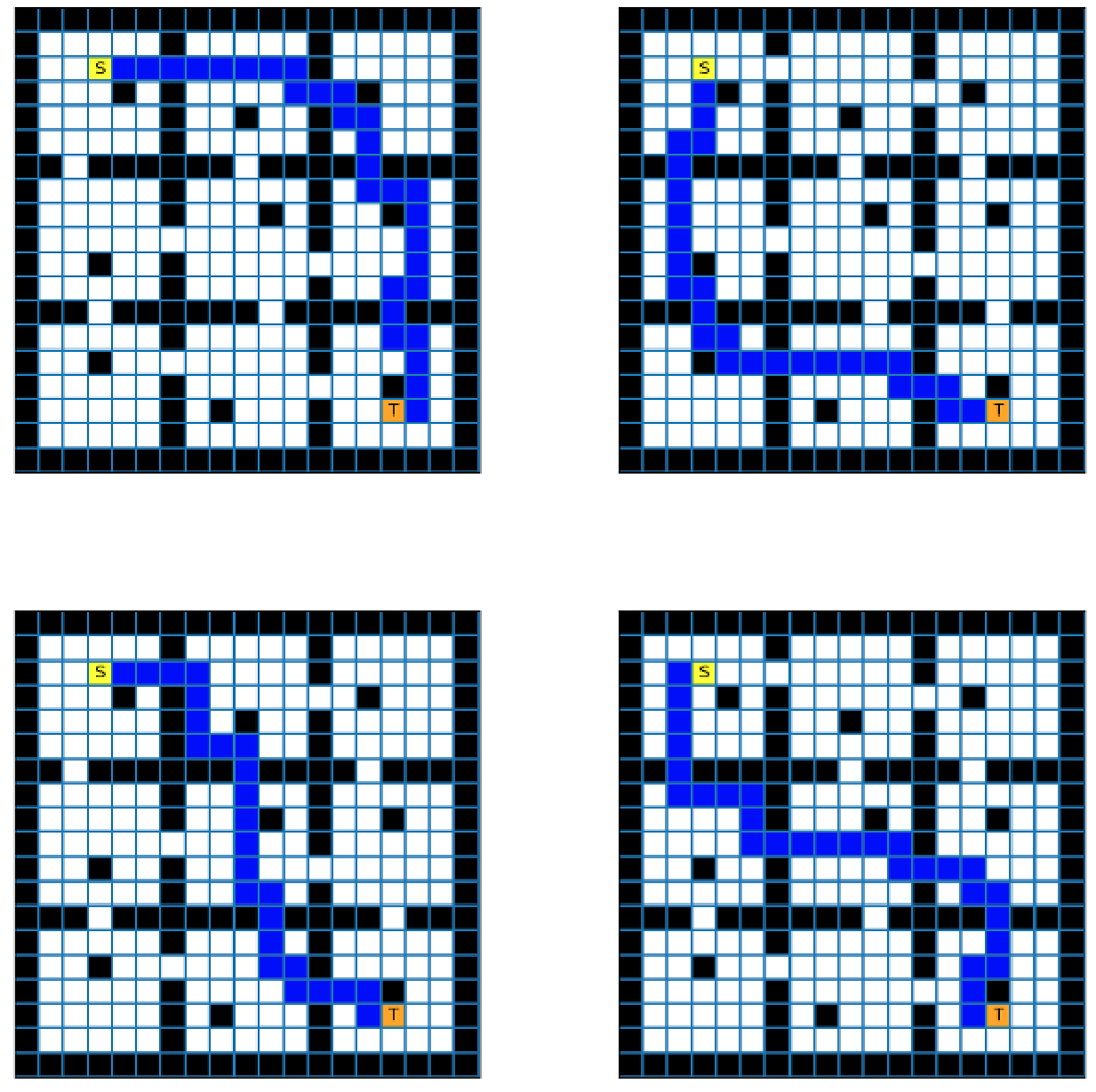}
        \par\small (b)
        \label{fig:b}
    \end{subfigure}
    \hspace{2mm}
    \begin{subfigure}{0.4\textwidth}
        \centering
        \includegraphics[width=150pt]{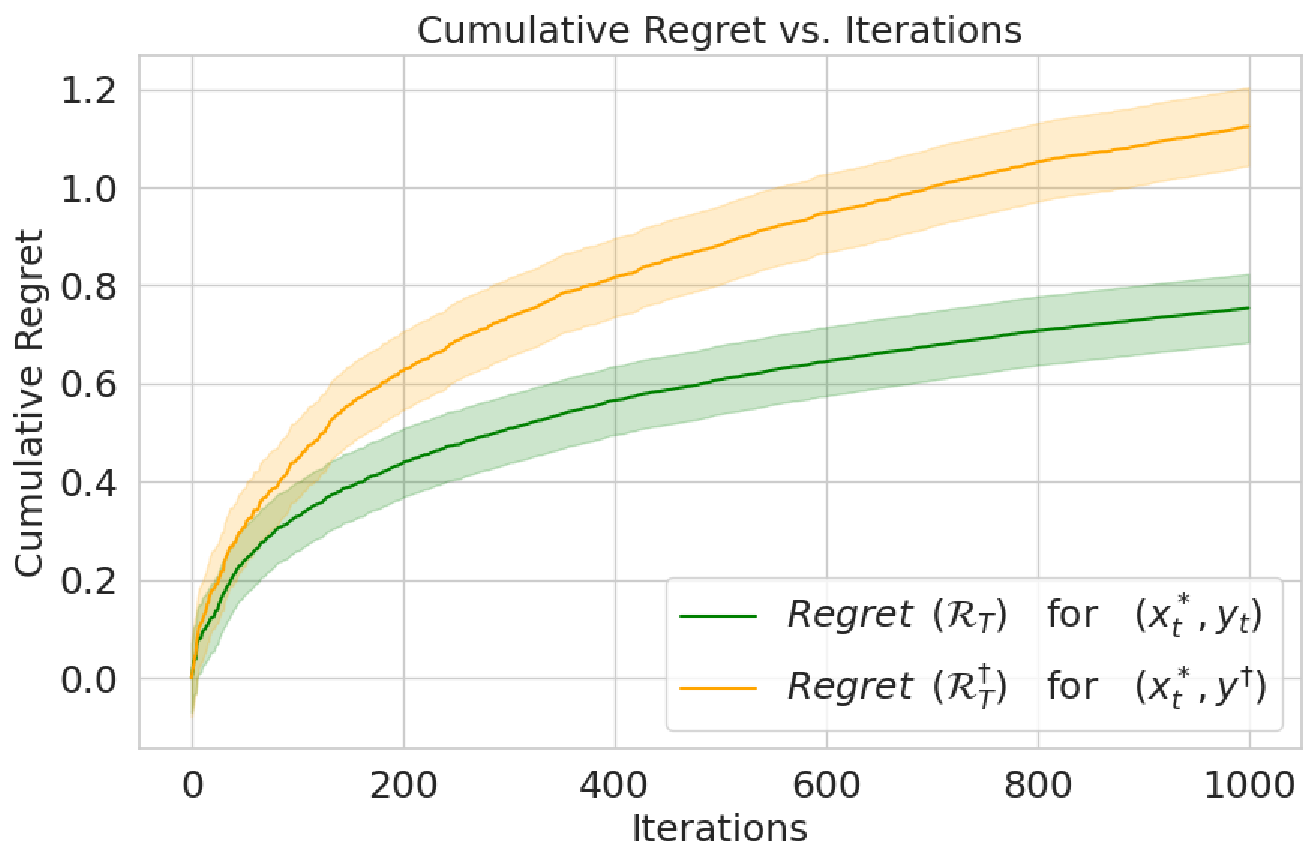}
        \par\small (c)
        \label{fig:c}
    \end{subfigure}
    \caption{(a) Sample sensor scheduling strategy (pure strategy) of the defender; (b) Sample paths (pure strategies) of the intruder in the 9-room grid world; (c) Cumulative regret plots for Alg. \ref{alg:1}}
    \label{fig:fig1}
\end{figure*}

 We consider the intrusion detection game under homogenous sensors, and evaluate the performance of Algorithm \ref{alg:1}. We set the true Bernoulli parameter to $p_{detect} = 0.8$ and generate several instances ($A$ matrices) by uniform random sampling of $V_{ij} \in [10,20], \forall (i,j) \in I \times J$, with $I=\{1,...,10\}$ and $J=\{1,...,20\}$. We solve for the NE strategies $(x^{\dagger},y^{\dagger})$ using Algorithm \ref{alg:dwm} with $\epsilon_T = 10^{-6}$ and estimate the optimal game value $V_A^* = x^{\dagger \top} A y^{\dagger}$. We run Algorithm \ref{alg:1} in both settings: (i) $y_t = y^{\dagger}$; and (ii) $y_t$ being a random point on the probability simplex $Y$ for each $t$. We plot the cumulative regrets $\mathcal{R}_T$ (for $y\ne y^{\dagger}$) and $\mathcal{R}_T^{\dagger}$ (for $y = y^{\dagger}$) by setting $T = 1000$ in Figure \ref{fig:fig1}(c). We observe that the regret grows sublinearly, indicating that the learning algorithm is effectively adapting and improving its performance over iterations. Additionally, we note that the regret is higher when the intruder employs the (optimal) Nash equilibrium strategy of the true game.

\section{Conclusions}
In this paper, we address the problem of sensor scheduling for intrusion detection by formulating it as a zero-sum game over a graph.  We proposed a distributed algorithm, which leverages the structure of the game's payoff, for efficiently computing equilibrium strategies for instances with exponentially large strategy spaces. We further extend our analysis to scenarios with unknown sensor models, introducing online learning algorithms with high-probability regret guarantees for learning equilibrium strategies under bandit feedback.  Our work leverages the structure in the combinatorial action space of the players to achieve significant computational efficiency and scalability. Empirical evaluations demonstrate the practical effectiveness of our approach, showing near-optimal and scalable performance over several instances of the game. In future work, we aim to extend the framework to dynamic game environments, where strategies can adapt to changes in sensor performance, including degradation or reconfiguration, as well as stochastic variations in game payoffs.
\newpage

\bibliographystyle{ieeetr} 
\bibliography{main}

\newpage

\appendix
\section{Online Learning Algorithm for Heterogeneous Sensors}
\label{app:ucb}
In this section, we present an adaptation of the Upper Confidence Bound (UCB) algorithm for matrix games originally studied in \cite{o2021matrix} for the intrusion detection game with heterogeneous sensors (see Algorithm \ref{alg:ucb}). The UCB algorithm and regret analysis in \cite{o2021matrix} is presented with respect to the max-player, however, in our game setting, the defender is the min-player. From Remark 3, we have the following:

\begin{align}
    (x^*, y^*) = \argmin_{x \in X} \argmax_{y\in Y} x^{\top} A y &= \argmax_{x \in X} \argmin_{y \in Y} x^{\top}(-A)y\\
    V^*_A &= - V^*_{-A}.
\end{align}
Therefore, without loss of generality, we will consider the game matrix to be $-A$ and note that the UCB estimate for analyzing the regret will remain the same, as the following relationship holds for the regret $\mathcal{R(\text{A},\text{UCB},\text{T})}$ as defined in \cite{o2021matrix}: $\mathcal{R(\text{A},\text{UCB},\text{T})} = - \mathcal{R}_T$, where $\mathcal{R}_T$ is the regret measure defined in our paper in \eqref{eq:regret}.

\begin{algorithm}[!h]

\caption{Online Learning and Adaptation of Strategies for Heterogeneous Sensors}\label{alg:ucb}
$\text{Feedback samples: }\mathcal{H} = \emptyset $\\
\For {$t=1,2,\ldots,T$}{
Compute sample average $\hat{p}_{detect,l}^t$ using samples in $\mathcal{H}$ \\
Clipping estimator: $\hat{p}_{detect,l}^t = \min (  \max(\hat{p}^t_{detect,l},p_{min,l}),p_{max,l}); l = 1,...p$\\
Compute $\displaystyle \tilde A^t_{ij} = \bar A^t_{ij} + \sum_{l=1}^{p} \frac{V_{max,l}}{1-p_{max,l}}\sqrt{\frac{\log(\frac{2}{\delta})}{2 (1 \vee n^t_{ij,l})}}$\\
Solve the estimated game $\tilde A^t$: \begin{equation} \label{eq:solve_game_1}
    \displaystyle  (x^*_t,y^*_t) = \argmax_{x\in X} \argmin_{y \in Y} x^{\top}\tilde A^ty\end{equation}
Play pure strategy $i_t \sim x^*_t$ and observe intruder's pure strategy $j_t \sim y_t$ \\
Collect samples from sensor feedback into $ \mathcal{H}_t$:  $\mathcal{H} \leftarrow \mathcal{H} \cup \mathcal{H}_t$ 
}

\end{algorithm}
The key distinction between Algorithm \ref{alg:ucb} and Algorithm 1 in \cite{o2021matrix} lies in how the UCB estimates for the game matrix entries are computed. In Algorithm \ref{alg:ucb}, these estimates are derived from the confidence bounds of 
$p$ Bernoulli parameters (representing the detection probabilities of the sensors), rather than maintaining separate estimates for each 
$ij$-entry of the matrix. This approach leverages the shrinking confidence bounds of the Bernoulli parameters to progressively tighten the UCB estimates of the game matrix, ultimately ensuring convergence to the true game payoffs.
\newpage
\section{Theoretical Proofs}
\label{app:proof}
\subsection{Proposition 2}
\begin{proof}
We have $ A'_{ij} = A_{ij} - r_j$ and  $B'_{ij} = B_{ij} - c_i$. By substitution, we have $A'_{ij} = \log(p_{miss}(i,j)) + c_i - r_j$ and $B'_{ij} = -\log(p_{miss}(i,j)) + r_j - c_i$. It follows that $A'_{ij} = - B'_{ij}$ and thus $A'_{ij} + B'_{ij} = 0$. Thus, the game $\mathcal{G}' = \{A',B' \}$ is a zero-sum game. The equivalence of Nash Equillbirum strategies for $\mathcal{G}$ and $\mathcal{G}'$ follows directly from the results presented in Section 2.1 in \cite{kannan2010games}.
\end{proof}

\subsection{Theorem 4}

\begin{proof}
    We establish this result by showing the equivalence between the multiplicative weight updates made by the DWM algorithm (Algorithm \ref{alg:1}) to that of the Weighted Majority algorithm in \cite{littlestone1994weighted}. Consider the per-agent loss function
    \begin{equation}
    \label{eq:agentloss_1}
    \ell^q_t(\theta_k) = \sum_{j \in J} A^q_{kj}y_t(j) - \frac{1} {p}\sum_{j=1}^{n} r_j y_t(j).
\end{equation}
Now, consider the joint strategy weight update rule given by 
\begin{align}
\label{eq:strat_con_1}
    w_{t+1}(i) = \prod_{q = 1}^{p} \sigma^q_{t+1}(k_{q,i}).
\end{align}
Substituting for $\sigma^q_{t+1}(k_{q,i})$, we have 
\begin{align}
    w_{t+1}(i) &= \prod_{q = 1}^{p} \sigma^q_{t}(k_{q,i}) \beta^{\ell^q_t(i_q)},\\
    &= \displaystyle \left(\prod_{q = 1}^{p} \sigma^q_{t}(k_{q,i}) \right) \beta^{\displaystyle (\sum_{q = 1}^p \ell^q_t(\theta_k))}
\end{align}
From definition of $A^q_{kj}$ in \eqref{eq:small_A}, we have 
\begin{align}
    \displaystyle \sum_{q=1}^p \ell^q_t(\theta_k) &= \sum_{q=1}^p \left( \sum_{j \in J} A^q_{kj}y_t(j) - \frac{1} {p}\sum_{j=1}^{n} r_j y_t(j) \right) \\
&= \sum_{q=1}^p \sum_{j \in J} \left( V^q_{kj} \times \log(1-p_{detect,q}) + c_k^q \right)y_t(j) - \sum_{j\in J} r_j y_t(j)\\
&= \sum_{j \in J}\left( \sum_{q=1}^p V^q_{kj} \times \log{(1-p_{detect,q})}  \right) y_t(j) + \sum_{j\in J}c_i y_t(j) - \sum_{j \in J} r_j y_t(j)\\
&= \sum_{j \in J} \left(\log(p_{miss}(i,j)) \right) y_t(j) + \sum_{j\in J}c_i y_t(j) - \sum_{j \in J} r_j y_t(j)\\
&=  \sum_{j \in J} \left(\log(p_{miss}(i,j)) + c_i - r_j \right)y_t(j)\\
&= \sum_{j \in J} A_{ij} y_t(j).
\end{align}

From Equations \eqref{eq:dist_weight} and \eqref{eq:strat_con} we have, $w_{t+1}(i) = w_t(i) \beta ^{\ell_t(i)}$, where $\ell_t(i) = \sum_{j \in J} A_{ij} y_j$. The weight update rule of the DWM algorithm (Algorithm \ref{alg:dwm}) is equivalent to the multiplicative weight update over the defender's joint strategy space when the Weighted Majority Algorithm (as in \cite{littlestone1994weighted}) is applied to compute the mixed strategies $x \in X$ for the defender. The guarantees presented in this theorem directly follow from the results in Section 2.4 and 2.5 in \cite{freund1996game}. We refer interested readers to \cite{littlestone1994weighted} and \cite{freund1996game} for detailed theoretical analysis of the weighted majority algorithm for zero-sum games.

\end{proof}
\subsection{Theorem 6}
\begin{proof}
    We begin by establishing a bound on $||\hat{A}_t - A||_\infty$. For each round of interaction $t$, the players play strategies $i_t,j_t$ and the defender receives $V_{i_t,j_t}$ feedbacks from the sensors, i.e., for each node $l$ traversed in the path $j_t$ and covered by the joint sensor strategy $i_t$, the sensors get a $0/1$ feedback whether the intruder was detected or not,  denoted by the Bernoulli random variable $Z^{t,l}_{ij}$. For round $t$, the estimate of Bernoulli parameter $\hat{p}^t_{detect}$ is computed as follows: 
    \begin{equation}
        \hat{p}^t_{detect} = \frac{1}{V_{i_t,j_t}} \sum_{l=1}^{V_{i_t,j_t}} Z^{t,l}_{ij}.
    \end{equation}
The sample average for $T$ rounds of interaction is given by 
\begin{equation}
    \hat{p}^{T}_{detect} = \frac{1}{T} \sum_{t=1}^{T} \left(\frac{1}{V_{i_t,j_t}} \sum_{l=1}^{V_{i_t,j_t}} Z^{t,l}_{ij} \right).
\end{equation}
   After $T$ rounds, the average number of samples obtained per round is given by 
   \begin{equation}
       \bar{k} = \frac{1}{T} \sum_{t=1}^{T}V_{i_t,j_t},
   \end{equation}
   and the total number of samples if $\bar{k}T$. Let $p_{detect}$ denote the true Bernoulli parameter, i.e., true probability detection of the sensors. Applying Hoeffding's inequality, we have 

   \begin{equation}
       \mathbb{P} \left( |\hat{p}^{T}_{detect} - p_{detect}| \ge \epsilon_T \right) \le 2 \exp{\left(-2\bar{k}T\epsilon_T^2\right)}.
   \end{equation}

   We define $\eta \ge 2 \exp{\left(-2\bar{k}T\epsilon_T^2\right)}$.
   For a specified $\epsilon_T$, with probability at least $1-\eta$, for $T> \frac{1}{(2\bar{k}\epsilon^2_T)} \log{\frac{2}{\eta}}$, we have $|\hat{p}^{T}_{detect} - p_{detect}| \le \epsilon_T$.

   We will now establish a bound for $||\hat{A}_t - A||_\infty$. By substituting for $A_{ij}$, we have,
   \begin{equation}
       |\hat{A}_{t,ij} - A_{ij}|= V_{ij} |\log(1-\hat{p}^t_{detect}) - \log(1-p_{detect})|.
   \end{equation}
   By applying the Mean Value Theorem to the function $\log(1-p)$, we have the following bound
   \begin{equation}
       |\log(1-\hat{p}^t_{detect}) - \log(1-p_{detect})| \le \frac{|\hat{p}^t_{detect} - p_{detect}|}{1-\max(\hat{p}^t_{detect},p_{detect})}.
   \end{equation}
    
   \noindent Since the true parameter $p_{detect} \in [p_{min}, p_{max}]$, clipping the estimates $\hat{p}^t_{detect}$ does not violate the Hoeffding's bound, we thus we have $1-\max(\hat{p}^t_{detect},p_{detect}) \ge 1-p_{max}$. Additionally, $V_{ij}\le V_{max}$, which leads to the following bound:
   \begin{equation}
       ||\hat{A}_t - A||_{\infty} \le \frac{V_{max}|\hat{p}^t_{detect} - p_{detect}| }{1-p_{max}}.
   \end{equation}
   Denote $\delta_t = \frac{V_{max}|\hat{p}^t_{detect} - p_{detect}| }{1-p_{max}}$. We have $\delta_t \le \frac{V_{max}\epsilon_t}{{(1-p_{max})}}$, with probability at least $1-\eta$.

\noindent Since with probability $1-\eta$, we have $\epsilon_t =  \sqrt{\frac{\log {(2/\eta)}}{2t}}$ (from Hoeffding's bound), we have 
\begin{equation}
\label{eq:a_bound}
    ||\hat{A}_t - A||_{\infty} \le \frac{V_{max}}{\sqrt{2t} (1-p_{max})} \sqrt{\log{\frac{2}{\eta}}}.
\end{equation}
We will now analyze the regret $\mathcal{R}_T$ as defined in \eqref{eq:regret}. Consider Case 1, where $y_t = y^{\dagger}$, i.e., the intruder knows the true game matrix $A$ and plays the NE strategy $y^{\dagger}$ of the true game. Consider the instantaneous regret $\mathcal{R}_t$ for round $t$:

\begin{equation}
    \mathcal{R}_t = x^{{* \top}}_tAy^{\dagger} - V^*_A= x^{{* \top}}_t Ay^{\dagger} - x^{\dagger \top} Ay^{\dagger}. 
\end{equation}
By adding and subtracting $x_t^*\hat{A}_ty^{\dagger}$, we have 

\begin{equation}
    \mathcal{R}_t =  \underbrace{x^{{* \top}}_t Ay^{\dagger} - x_t^{{* \top}}\hat{A}_ty^{\dagger}}_\text{\clap{Term-I~}}  + \underbrace{ x_t^{{* \top}}\hat{A}_ty^{\dagger} - x^{\dagger \top}Ay^{\dagger}}_\text{\clap{Term-II~}}. 
\end{equation}

Term-I:

\begin{equation}
\label{eq:bound_1}
    x^{{* \top}}_tAy^{\dagger} - x_t^{{* \top}}\hat{A}_ty^{\dagger} = x^{{* \top}}_t\left( A - \hat{A}_t \right) y^{\dagger} \le ||\hat{A}_t - A||_{\infty} \le \delta_t.
\end{equation}
To Term-II, we add and subtract $x^{\dagger \top}\hat{A}_t y^{\dagger}$, and obtain the following:

\begin{equation}
    \text{Term-II} =  \underbrace{x_t^{{* \top}}\hat{A}_ty^{\dagger} - x^{\dagger \top}\hat{A}_t y^{\dagger}}_\text{\clap{Term-II-I~}}  + \underbrace{ x^{\dagger \top}\hat{A}_t y^{\dagger} - x^{\dagger \top}Ay^{\dagger}}_\text{\clap{Term-II-II~}} 
\end{equation}
  Now consider Term-II-I. We know that $(x^*_t,y^*_t)$ is the NE for $\hat{A}_t$. Since the y-player is maximizing the expected game value, we have $x^*_t\hat{A}_ty^*_t \ge x^*_t\hat{A}_ty, \forall y \in Y$ and thus we have $x^*_t\hat{A}_ty^{\dagger} \le x^*_t\hat{A}_ty^*_t$. As a result, we have 
  \begin{equation}
      \text{Term-II-I} \le x^*_t\hat{A}_ty^*_t - x^{\dagger \top}\hat{A}_t y^{\dagger}.
  \end{equation}

The right-hand side of the above expression can be viewed as the difference in the game values when the NE strategies of a game $A$, i.e., $(x^{\dagger},y^{\dagger})$, are played in a perturbed game $\hat{A}_t$. From Theorem 4 in \cite{lipton2006stability}, we have the following:

\begin{equation}
\label{eq:bound_2}
    \text{Term-II-I} \le 3 ||\hat{A}_t - A||_{\infty} = 3 \delta_t.
\end{equation}
Term-II-II 

\begin{equation}
\label{eq:bound_3}
    x^{\dagger \top}\hat{A}_t y^{\dagger} - x^{\dagger \top}Ay^{\dagger}  = x^{\dagger \top}(\hat{A}_t-A) y^{\dagger} \le ||\hat{A}_t - A||_{\infty} = \delta_t.
\end{equation}

Combining Equations \eqref{eq:bound_1}, \eqref{eq:bound_2} and \eqref{eq:bound_3}, we have the following bound on the instantaneous regret $\mathcal{R}_t$:

\begin{equation}
    \mathcal{R}_t \le \delta_t + 3 \delta_t + \delta_t = 5 \delta_t.
\end{equation}

The cumulative regret can be bounded as follows. With probability at least $1-\alpha$,
\begin{align}
    \mathcal{R}_T &= \sum_{t=1}^{T} \mathcal{R}_t \le \sum_{t=1}^{T} 5 \delta_t\\
     &\le \frac{5 V_{max}}{\sqrt{2} (1-p_{max})} \sqrt{\log{\frac{2}{\eta}}} \sum_{t=1}^T \frac{1}{\sqrt{t}}\\
    &\le \frac{5 V_{max}}{\sqrt{2} (1-p_{max})} \sqrt{\log{\frac{2}{\eta}}} \int_{t=0}^{T} \frac{1}{\sqrt{t}}\\
    &\le \frac{5 V_{max}}{\sqrt{2} (1-p_{max})} \sqrt{\log{\frac{2}{\eta}}} \times 2\sqrt{T}\\
    & \le \frac{5\sqrt{2} V_{max}}{(1-p_{max})} \sqrt{T \log{\frac{2T}{\alpha}}},
\end{align}
where $\alpha = T \eta$.
This completes the proof of Part 1. 

Now consider the scenario where $y_t \ne y^{\dagger}$, i.e., when the defender is playing against a suboptimal opponent (intruder), who does not play the NE strategy of the true game. The instantaneous regret is given by 
\begin{equation}
    \mathcal{R}_t = x^{{* \top}}_tAy_t - V^*_A= x^{{* \top}}_t Ay_t - x^{\dagger \top} Ay^{\dagger}.
\end{equation}
By adding and subtracting $x^{{* \top}}_t\hat{A}_ty_t$, we have 
\begin{equation}
    \mathcal{R}_t =  \underbrace{x^{{* \top}}_t Ay_t - x_t^{{* \top}}\hat{A}_ty_t}_\text{\clap{Term-I~}}  + \underbrace{x_t^{{* \top}}\hat{A}_ty_t  - x^{\dagger \top}Ay^{\dagger}}_\text{\clap{Term-II~}}.
\end{equation}

Term-I: 
\begin{equation}
\label{eq:bound11}
    x^{{* \top}}_t Ay_t - x_t^{{* \top}}\hat{A}_ty_t = x^{{* \top}} \left( A -\hat{A}_t \right) y_t \le ||\hat{A}_t - A ||_{\infty} = \delta_t.
\end{equation}
In Term-II, we have $x^{{* \top}}_t\hat{A}_ty_t \le x^{{* \top}}_t\hat{A}_ty^*_t$, since the y-player is maximizing the expected game value, and thus
\begin{equation}
    x_t^{{* \top}}\hat{A}_ty_t  - x^{\dagger \top}Ay^{\dagger} \le x^{{* \top}}_t\hat{A}_ty^*_t - x^{\dagger \top}Ay^{\dagger}.
\end{equation}

From Theorem 5 in \cite{lipton2006stability}, we have that $x^{{* \top}}_t\hat{A}_ty^*_t - x^{\dagger \top}Ay^{\dagger} < ||\hat{A}_t - A||_\infty$, and thus

\begin{equation}
\label{eq:bound22}
    x_t^{{* \top}}\hat{A}_ty_t  - x^{\dagger \top}Ay^{\dagger} \le ||\hat{A}_t - A||_\infty = \delta_t.
\end{equation}

Combining the terms in Equations \eqref{eq:bound11} and \eqref{eq:bound22}, we have 
\begin{equation}
    \mathcal{R}_t \le \delta_t  + \delta_t = 2 \delta_t.
\end{equation}
Using similar arguments as in Case 1, we have with probability at least $1-\alpha$,
\begin{equation}
    \mathcal{R}_T \le \frac{2\sqrt{2} V_{max}}{(1-p_{max})} \sqrt{T \log{\frac{2T}{\alpha}}}.
\end{equation}
\end{proof}
\subsection{Lemma 8}

\begin{proof}
    Consider the following expression for $A_{ij}$:
    \begin{equation}
    \label{eq:hetero_A_1}
    A_{ij} = \sum_{l=1}^{p} V_{ij,l} \log(1-p_{detect,l}) +c_i - r_j.
\end{equation}

We have the following confidence bound estimate for $p_{detect,l}$ which follows from Hoeffding inequality (alternatively a special case of Chernoff's bound): With probability at least $1-\delta$

\begin{equation}
    p_{detect,l} \le \hat{p}^t_{detect,l} + \sqrt{\frac{\log{\frac{2}{\delta}}}{2(1 \vee n^t_{ij,l})}} = \Tilde{p}^t_l.
\end{equation}

Using the similar technique as in proof of Theorem \ref{thm:regret}, we apply the Mean Value Theorem to the function $A_{ij}({p_{detect,l}})$  with respect to 
$p_{detect,l}, l = 1,...,p$  ( as in \eqref{eq:hetero_A_1}) to obtain the following UCB estimate for $A_{ij}$
\begin{equation}
\label{eq:UCB_A_1}
    A_{ij} \leq \bar{A}^t_{ij} + \sum_{l=1}^{p} \frac{V_{max,l}}{1-p_{max,l}} \sqrt{\frac{\log(\frac{2}{\delta})}{2 (1 \vee n^t_{ij,l})}} = \Tilde{A}^t_{ij},
\end{equation}
where $V_{max,l} = \max_{ij}V_{ij,l}$.
\end{proof}

\subsection{Theorem 9}

\begin{proof}
    We establish this proof by making similar arguments as in Theorem 1 of \cite{o2021matrix}. Note that we discuss the equivalence of our regret definition and problem setting with the one presented in \cite{o2021matrix} in Appendix \ref{app:ucb}. 

    Since the upper confidence matrix \eqref{eq:UCB_A_1}  overestimates the true matrix, we have $V^*_{\Tilde{A}_t} \ge V^*_A$ (see Proof of Theorem 1 in \cite{o2021matrix}). Let $E_t$ be the bad event that there exists a pair $i,j$ such that $p_{detect,l} > \Tilde{p}^t_{l}$ for $l = 1,...,p$. By definition, $E_t \in \mathcal{F}_t$. Consider that $E_t$ does not hold and let $x^*_t$ be the best-response of the x-player at round $t$. We have the following for the instantaneous regret $\mathcal{R}_t$: 
    \begin{align}
       \mathcal{R}_t =    (V^*_A - x_t^{\top}Ay_t ) &\leq   (V^*_{\Tilde{A}_t} - x_t^{\top}Ay_t);\\
       & =   (x^{* \top}_t\Tilde{A}_ty_t - x_t^{\top}Ay_t);\\
       & \le   (x_t^{\top} (\Tilde{A}_t - A)y_t);\\
       &\le  \left[\sum_{l=1}^{p} \frac{V_{max,l}}{1-p_{max,l}} \sqrt{\frac{\log(\frac{2}{\delta})}{2 (1 \vee n^t_{ij,l})}}\right],
    \end{align}
    where the last step is obtained by substituting the upper confidence bound expression as in \eqref{eq:UCB_A_1}.  
    The cumulative regret can be decomposed into the sum of regrets due to the good and bad events, and is given by 
    \begin{align}
        \mathcal{R}_T &= \sum_{t=1}^{T} \mathcal{R}_t;\\
        & \le   \left[ \sum_{t=1}^{T} \sum_{l=1}^{p} \frac{V_{max,l}}{1-p_{max,l}} \sqrt{\frac{\log(\frac{2}{\delta})}{2 (1 \vee n^t_{ij,l})}} \right] + T \mathbb{P}(\cup_{t=1}^{T} E_t),
    \end{align}
with probability at least $1-\delta$.
   \noindent Using similar arguments as in Theorem 1 in \cite{o2021matrix}, we bound the second term  as follows: $T \mathbb{P}(\cup_{t=1}^{T} E_t) \le 2 T^2 |\mathcal{S}||\Theta| |J| \delta \leq 1$. Recall that $\bar{V}_{max} = \max_{l} V_{max,l}$, $\bar{p}_{max} = \max_{l} p_{max,l}$ and $C = \bar{V}_{max}/(1-\bar{p}_{max}) $.
    
    \noindent The first term is bounded by 
    \begin{align}
          \left[ \sum_{t=1}^{T} \sum_{l=1}^{p} \frac{V_{max,l}}{1-p_{max,l}} \sqrt{\frac{\log(\frac{2}{\delta})}{2 (1 \vee n^t_{ij,l})}} \right] &\le \frac{\bar{V}_{max}}{1-\bar{p}_{max}} \sum_{ij}   \sum_{l=1}^{p} \sqrt{2 n^T_{ij,l} \log{\frac{2}{\delta}}};\\
        & \le C|\mathcal{S}| \sqrt{2  |\Theta| |J| T\log{\frac{2}{\delta}}}.
    \end{align}
    By substituting with $\delta = 1/2T^2 |\mathcal{S}||\Theta||J|$, we have with probability at least $1-\delta$,
    \begin{equation}
        \mathcal{R}_T \leq 1 + C|\mathcal{S}|\sqrt{2 |\Theta||J|T \log(4T^2|\mathcal{S}||\Theta||J|)} = \Tilde{O}(\sqrt{|\mathcal{S}|^2|\Theta||J|T}).
    \end{equation}
\end{proof}

\end{document}